\documentclass[11pt]{article}
\usepackage[letterpaper,margin=1in]{geometry}
\usepackage[T1]{fontenc}
\usepackage{lmodern}
\usepackage{amsmath,amssymb,amsthm,mathtools}
\usepackage{booktabs,array,graphicx}
\usepackage{tikz}
\usetikzlibrary{positioning,fit,arrows.meta,calc}
\usepackage[colorlinks=true,linkcolor=blue!45!black,citecolor=blue!45!black,urlcolor=blue!45!black]{hyperref}
\newtheorem{theorem}{Theorem}[section]
\newtheorem{lemma}[theorem]{Lemma}
\newtheorem{proposition}[theorem]{Proposition}
\newtheorem{corollary}[theorem]{Corollary}

\theoremstyle{remark}

\newcommand{\cost}{\operatorname{cost}}

\newcommand{\OO}{\widetilde O}

\title{Tight Quantum Query Bounds for Minimum Cut\\and Multicut on Dense Graphs}
\author{Tianhang Lu$^{\ast}$\qquad Runtian Ren$^{\ast}$\qquad Shengcai Liu\\[0.65em]
{\small Guangdong Provincial Key Laboratory of Brain-Inspired Intelligent Computation}\\
{\small Department of Computer Science and Engineering}\\
{\small Southern University of Science and Technology, Shenzhen 518055, China}\\
{\small \texttt{liusc3@sustech.edu.cn}}\\[0.3em]
{\footnotesize $^{\ast}$Equal contribution.}}
\date{}
\hypersetup{pdftitle={Tight Quantum Query Bounds for Minimum Cut and Multicut on Dense Graphs},
  pdfauthor={Tianhang Lu, Runtian Ren, Shengcai Liu}}
\begin{document}
\maketitle

\begin{abstract}
We determine the dense worst-case quantum query complexity of exact
minimum multicut on simple unweighted undirected graphs. With at most
$k$ explicit demand pairs, our adjacency-array algorithm uses
$\OO(n+\min\{m,n^{3/2}\sqrt{k}\})$ queries; a matching lower bound
holds for every $1\le k\le\lfloor n/2\rfloor$ and sufficiently large
$n$, with constants uniform in $k$. In particular, the $k=1$ case gives
an $\OO(n^{3/2})$ query algorithm for prescribed $s$--$t$ cut.
The upper bound combines degree-sensitive spectral estimates of local
improvement with refinement by unions of label classes. These steps
retain every original optimum and permit exact acquisition without an
additional label-count factor. Separately, we prove an $\Omega(n^{3/2})$
global minimum-cut lower bound on dense connected graphs, matching the
existing upper bound up to polylogarithmic factors. It holds for the
value and for any optimum shore, even with known degrees and sorted
neighbor arrays; regular hard instances have balanced optimum cuts.
Both lower bounds use degree-preserving switches and classify optimum
outputs across a hidden threshold. Our upper bound concerns input
queries, not total running time: computation on acquired data may be
exponential, even in the single-demand case. Matching bounds at each
prescribed sparse edge count are not established.
\end{abstract}

\section{Introduction}

How much of a graph must an algorithm inspect to find an exact optimum
cut? We study this question for simple unweighted undirected graphs in
the quantum adjacency-array model. Global minimum cut minimizes a
nontrivial edge boundary with no prescribed terminals. Minimum multicut
instead asks for a smallest edge set whose removal disconnects every
pair in an explicit demand set. One demand gives minimum $s$--$t$ cut;
the three pairwise demands among three terminals give three-terminal
multiway cut. These are distinct separation problems, but in each case
the output is one exact optimum, rather than a description of the graph.

For prescribed demands, we give tight dense worst-case query bounds
throughout the range from one demand to a linear demand budget. The
upper bound uses approximate graph information to construct local
stability certificates, then acquires enough original edges to recover
an exact optimum. Its $k=1$ case improves the dense $s$--$t$ cut upper
bound of Apers--Auza--Lee~\cite{AAL24} from $\OO(n^{11/6})$ to
$\OO(n^{3/2})$. Separately, our global-cut lower bound matches the
dense upper bound of Apers--Lee~\cite{AL21} and rules out their proposed
near-linear unweighted adjacency-array query bound. Table~\ref{tab:comparison}
summarizes the comparison in a common model.

\begin{table}[htb]
\centering
\small
\renewcommand{\arraystretch}{1.18}
\begin{tabular}{@{}>{\raggedright\arraybackslash}p{0.20\linewidth}>{\raggedright\arraybackslash}p{0.35\linewidth}>{\raggedright\arraybackslash}p{0.37\linewidth}@{}}
\toprule
Problem & Earlier query bounds or baseline & Bounds after this work \\
\midrule
Global minimum cut
& $\OO(n^{3/2})$ upper; $\Omega(n)$ lower in Apers--Lee~\cite{AL21}
& $\widetilde\Theta(n^{3/2})$; the lower bound is new \\
Prescribed $s$--$t$ cut
& $\OO(n^{11/6})$ upper, Apers--Auza--Lee~\cite{AAL24}
& $\widetilde\Theta(n^{3/2})$; new upper and lower bounds \\
Multicut, at most $k$ demands
& $O(n^2)$ by full acquisition and offline optimization (baseline)
& $\widetilde\Theta(n^{3/2}\sqrt{k})$; new upper and lower bounds \\
\bottomrule
\end{tabular}
\caption{Dense worst-case quantum adjacency-array \emph{query} bounds
for exact optimization on simple unweighted undirected graphs. Outputs
are an optimum value or a vertex partition/shore, with bounded error.
The multicut row uses $1\le k\le\lfloor n/2\rfloor$ and sufficiently
large $n$; the read-all entry is a baseline, not a literature lower
bound. No row asserts a total-time bound for our algorithm.}
\label{tab:comparison}
\end{table}

\subsection{Our results}

We first state the uniform multicut theorem, followed by the global-cut
lower bound. Both use the ordinary adjacency-array oracle, with degree
queries charged. Weighted sparsifiers and quotients occur only as
auxiliary objects; the input graphs have unit edge weights.

\begin{theorem}[Multicut]\label{thm:intro-main}
For every $n\ge2$ and demand budget $k\ge1$, exact minimum multicut
on a simple unweighted undirected graph with $n$ vertices, $m$ edges,
and at most $k$ explicit demands admits an algorithm using
\begin{equation}\label{eq:intro-upper}
 \OO\!\left(n+\min\{m,n^{3/2}\sqrt{k}\}\right)
\end{equation}
quantum adjacency-array queries. It returns the optimum value and an
optimal partition with probability at least $2/3$.

Conversely, for every $n\ge5120$ and
$1\le k\le\lfloor n/2\rfloor$, there is a connected dense family
with at most $k$ pairwise vertex-disjoint demands on which either
output task requires
\begin{equation}\label{eq:intro-lower}
 \Omega(n^{3/2}\sqrt{k})
\end{equation}
queries. The same lower bound holds for an optimal edge-removal set.
Degrees, edge count, and demands are input-independent within the
family and may be supplied for free. All constants are independent of $k$.
\end{theorem}

Theorems~\ref{up:main} and~\ref{lo:main} prove the two bounds. Hence
the dense worst-case multicut complexity is
\begin{equation}\label{eq:dense-tight}
 \widetilde\Theta(n^{3/2}\sqrt{k})
 \qquad\text{for }1\le k\le\lfloor n/2\rfloor.
\end{equation}
The $k=1$ case is prescribed $s$--$t$ cut, not global cut. Both have
the same dense worst-case order, established by the two separate
theorems. Every fixed demand budget, including three-terminal multiway
cut, has an $\OO(n^{3/2})$ upper bound. For $k=n^\alpha$ with fixed
$0<\alpha<1$, the bound becomes
$\widetilde\Theta(n^{3/2+\alpha/2})$; a linear demand budget requires
$\Omega(n^2)$ queries even when the demands are disjoint.

\begin{theorem}[Global minimum cut]\label{thm:intro-global}
For every $n\ge33$, there is a connected simple unweighted family on
exactly $n$ vertices and $\Theta(n^2)$ edges on which either computing
the global minimum-cut value or outputting any minimum-cut shore
requires $\Omega(n^{3/2})$ quantum queries. The degrees and edge count
are input-independent and may be given for free. The lower bound holds
even with sorted neighbor arrays.
\end{theorem}

Theorem~\ref{glob:main} proves this statement. With the existing
global-cut upper bound, it determines dense worst-case query complexity
as $\widetilde\Theta(n^{3/2})$. We also give regular hard instances
whose unique optimum has two balanced, non-singleton shores
(Proposition~\ref{glob:regular}). This strengthening has its own
vertex-count grid and is not needed for the all-$n$ theorem.

\paragraph{Query complexity and computation.}
Our upper bound measures calls to the input graph oracle. Computation
on acquired data is unrestricted and may be exponential: in particular,
we enumerate candidate labelings, compute their common refinement, and
optimize an explicit quotient. This limitation also applies to our
$k=1$ implementation. The improvement over~\cite{AAL24} is therefore
an improvement in queries, not in total running time. The revised
version of~\cite[Remark 9]{AAL24} likewise establishes query efficiency
without the earlier claimed time bound. In contrast, Apers--Lee give
both query and time bounds for global cut~\cite[Theorem 5]{AL21}; we
use their existing algorithm only for the global-cut comparison.
The question resolved here is the amount of input access required for
exact optimization, allowing arbitrary finite computation between queries.

\paragraph{Scope.}
Our multicut upper bound applies at every edge density; matching lower
bounds concern dense worst-case instances, not each prescribed edge
count $m$. The parameter $k$ is a budget of at most $k$ demand pairs.
The sorted-array and regularity promises in the global lower bound are
not asserted for multicut. Weighted-input results and counting all
minimum cuts are outside this paper's scope.

\subsection{The two ingredients of the upper bound}

The sparsify--contract--learn framework is inherited from
RSW~\cite{RSW18} and Apers--Lee~\cite{AL21}: retain the unknown optimum
in an explicit candidate family, contract vertices that every candidate
keeps together, and acquire the remaining original edges. Their
global-cut covering lemmas concern near-minimum nontrivial bipartitions,
with error measured against the global minimum degree. Applying such a
lemma to the shores of a multicut labeling is not immediate: those
shores need not be near-minimum global cuts. We instead construct
candidates with degree-sensitive local stability and prove an acquisition
bound for their common refinement. These are the roles of
Lemmas~\ref{up:candidates} and~\ref{up:acquisition}, respectively.

\paragraph{Local certificates from degree layers (Lemma~\ref{up:candidates}).}
Let $K=k(n-1)$, an upper bound on the optimum, and protect the demand
endpoints. Changing the label of any other vertex preserves feasibility,
so an original optimum admits no improving such move. A multiplicative
approximation to the objective alone controls differences of two costs
of order $K$ only to additive error $O(\varepsilon K)$. This can exceed
the degree of a vertex whose move we must test. Thus retaining all
near-optimal candidates does not by itself provide the local certificate
needed for acquisition.

We group edges by the smaller degree of their endpoints. A degree-$h$
layer has a domain of $N_h=O(nh)$ padded edge records. With spectral
accuracy $\varepsilon_h=\Theta(a\sqrt{h/K})$, for a fixed small constant
$a$, a bilinear estimate controls the error in a move at $v$ within
that layer by
\[
 \varepsilon_h\bigl(2\sqrt{d_h(v)C_h(f)}+d_h(v)\bigr),
\]
where $d_h(v)$ and $C_h(f)$ are its degree and the labeling cost in
that layer. Summing over layers gives
\[
 |\operatorname{gain}_H(f,v,b)-\operatorname{gain}_G(f,v,b)|
 \le a\,d(v)\left(2\sqrt{\cost_G(f)/K}+1\right).
\]
The estimate holds simultaneously for all labelings and moves, without
a union bound over them. Explicit cost and local-gain tests therefore
retain every original optimum while ensuring that every retained
candidate has cost at most $2K$ and every nonterminal move has gain
less than $d(v)/4$. The query cost per layer is
$\OO(\sqrt{N_hn}/\varepsilon_h)=\OO(n\sqrt K)$.
The new ingredient is this construction of local certificates at the
right degree scales, not the spectral sandwich or bilinear inequality
in isolation. Padded records implement the layers through ordinary
neighbor queries; no filtered-neighbor oracle is assumed.

\paragraph{Acquisition without a label-count loss (Lemma~\ref{up:acquisition}).}
Balanced-shore selection and charging exceptional vertices to a selected
boundary build on RSW~\cite[Lemma 2.6 and Section 2.3]{RSW18} and
Apers--Lee~\cite[Lemma 2 and Section 3]{AL21}. The additional issue for
multicut is that a candidate may have many label classes. If one tests
only individual classes, every class in a large current part may be
too small to give a balanced split. The absence of such a split then
does not imply that any label dominates the part.

We allow a selected shore to be a union of classes of \emph{one}
candidate labeling. Its boundary is still contained in that labeling's
crossing edges, so it has at most $2K$ edges. At degree scale $h$, select
such shores whenever they split a current part into two pieces of size
at least $s=h/8$. There are $O(n/h)$ selections. At termination, a part
of size at least $3s$ has, for each candidate, one label containing all
but fewer than $s$ of its vertices: otherwise accumulating small classes
would give another permitted split.

A nonterminal outside this dominant class can move directly to it.
Local stability forces a constant fraction of its degree to cross the
selected boundary $U_h$, giving $d(v)\le4\deg_{U_h}(v)$.
Different vertices may use different candidate witnesses, but they all
charge the same $U_h$, whose size is $O(nK/h)$. Thus the cost of reading
their original arrays has no additional label-count factor. Small parts
admit the same charge by simplicity of the graph. All other required
layer edges cross the selected partition and can be enumerated in
\[
 \OO\!\left(\sqrt{(nh)(nK/h)}\right)=\OO(n\sqrt K)
\]
queries. Low-degree arrays are read directly; dyadic summation bounds
all exceptional-array reads. This turns the local certificate into a
uniform acquisition bound. It does not require the union shores to be
near-minimum global cuts, or to be feasible multicuts themselves.

\subsection{Lower bounds and related work}

Both lower bounds use a four-clique core with input-independent degrees
and switch slots that depend on one hidden bit each. A fixed attachment
sets the threshold for global cut; selectors carrying disjoint demands
set a threshold proportional to the demand count for multicut. Crossing
the threshold changes the type of every optimum solution. The multicut
proof classifies all optimal partitions, including those with more than
two parts, so a returned partition alone distinguishes the inputs.
A common Hamming-layer adversary bound gives $\Omega(n^{3/2})$ and
$\Omega(n^{3/2}\sqrt{k})$, respectively. The lower-bound tools are
standard~\cite{Ambainis02,NayakWu99}, and degree-preserving switches
build on graph-query constructions~\cite{AL21,BGMP21}. The shared
core, attachments, output classifications and coherent simulations are
proved here. Global cut is not the one-demand special case of multicut.

The $s$--$t$ algorithm of Apers--Auza--Lee~\cite{AAL24} follows the
RSW sparsifier/flow/strong-component approach. Our upper proof instead
uses the two local-stability and refinement lemmas above, which apply
uniformly to multiple demands. RSW's cut-query bounds~\cite{RSW18}
are not entries in Table~\ref{tab:comparison}: one cut query reports
an entire boundary size, whereas one adjacency query reports a degree
or a single neighbor. Likewise, global $k$-cut~\cite{Li19} prescribes
a number of components, rather than separation demands.

Multicut-mimicking networks~\cite{Wahlstrom22,ALLS24} preserve terminal
separation values, but explicit-input compression does not by itself
bound the queries needed to acquire a contraction. Lossless contraction
in exponential-time approximation schemes~\cite{IKPRS24} has a
different guarantee. Ambainis--\v{S}palek~\cite[Theorem 7]{AS06}
prove quadratic query hardness for exact network-flow values with
large capacities; our lower bounds use unit weights and cover optimum
partition outputs. The edge-record implementation uses repeated-halving
sampling~\cite{Cohen14}, matrix concentration~\cite{Tropp12}, and
quantum enumeration~\cite{AL21}; these underlying tools are not new.

Section~\ref{sec:model} fixes the model. Section~\ref{sec:upper} gives
the upper bound, including the two core lemmas and their complete proofs.
Section~\ref{sec:lower-tools} supplies the common lower-bound tools;
Sections~\ref{sec:global-lower} and~\ref{sec:lower} prove the global
and multicut lower bounds. Section~\ref{sec:discussion} discusses the
remaining questions. Appendix~\ref{rec:section} implements spectral
sparsification on padded edge records.

\section{Model and preliminaries}\label{sec:model}
Let $G=(V,E)$ be a simple unweighted undirected graph, $V=[n]$.
For $n\ge2$, a global cut is an unordered bipartition
$\{S,V\setminus S\}$ with $\emptyset\ne S\ne V$. Write
\[
 \delta_G(S)=\{uv\in E:u\in S,\ v\notin S\},\qquad
 \lambda(G)=\min_{\emptyset\ne S\ne V}|\delta_G(S)|.
\]
The global output tasks are the value $\lambda(G)$ and any shore attaining
it, considered separately. Complementary shores describe the same cut.
Disconnected graphs are allowed in the problem definition; the global
hard families below are connected.

\paragraph{Prescribed demands.}
For multicut, the explicit demand set contains at most $k$ unordered pairs of distinct
vertices; duplicates can be removed. A partition is feasible if no part
contains both endpoints of a demand. Its cost counts edges crossing its parts.
Empty demands have optimum zero and permit the one-part output without queries.

For nonempty demands let $T$ be the endpoint set and
$r=t=|T|\le\min\{2k,n\}$. Use labelings $f:V\to[r]$, allowing unused labels.
Feasibility requires different labels only on demand pairs. The labeling
optimum equals the partition optimum: merge endpoint-free parts into any
endpoint-containing part, without increasing cost or violating a demand.
Write
\[
 \cost_G(f)=|\{uv\in E:f(u)\ne f(v)\}|,\qquad
 C=\min_{f\text{ feasible}}\cost_G(f),\qquad K=k(n-1).
\]
Isolating one chosen endpoint per demand gives $C\le K$; also
$d(v)\le n-1\le K$. Nonempty demands imply $n\ge2$.

\paragraph{Oracle and output.}
Each vertex has an arbitrary fixed neighbor array without repetitions.
Degree and neighbor queries are coherent XOR-output queries; invalid slots
return a fixed symbol $\bot$. Both query types are charged. Orders need not
be known or sorted. After $n$ degree queries, prefix sums index all $2m$
adjacency slots, and full reading costs $O(n+m)$.

Outputs are classical, with success probability at least $2/3$ on every input.
Exactness refers to the optimum, not to zero-error query complexity. Known-data
computation, processing demands, exact rational linear algebra, random-table
generation, exhaustive labeling searches, and reversible explicit-table lookup
are free. These computations are finite but can use exponential time and
memory. No gate-efficiency or efficient QRAM claim is made.

\paragraph{Spectral approximation and gains.}
For a nonnegatively weighted graph $J$, let $L_J$ be its Laplacian.
The sandwich $(1-\varepsilon)L_G\preceq L_J\preceq(1+\varepsilon)L_G$
preserves every labeling cost by the same factors, since
\[
 \cost_J(f)=\frac12\sum_{i=1}^r
 \mathbf1_{f^{-1}(i)}^\top L_J\mathbf1_{f^{-1}(i)}.
\]
For a nonterminal $v$ and another label $b$, let $f^{v\to b}$ be the feasible
single-vertex change and define
\[
 \operatorname{gain}_J(f,v,b)=\cost_J(f)-\cost_J(f^{v\to b}).
\]
An original optimum has no positive nonterminal gain.

\paragraph{Capped enumeration.}
For $N$ known positions and at most $B$ marks, quantum enumeration learns all
marks in $\OO(\sqrt{N(B+1)})$ predicate queries, with full reading as an
alternative~\cite[Theorem 13]{AL21}. Its query and output budgets can be fixed
even when the mark bound fails. Inverse-polylogarithmic failure probabilities
cost polylogarithmic overhead. The additive one covers zero marks.

\paragraph{Refinements and contraction.}
A common refinement keeps vertices together exactly when every constituent
partition does. Write $E_G(\mathcal A)$ for its crossing edges. After acquiring
these original edges, contract the parts and aggregate parallel edges offline.
This creates an explicit weighted quotient, without a contracted-graph oracle.

\section{A uniform upper bound}\label{sec:upper}

\begin{theorem}\label{up:main}
Let $n\ge2$ and $k\ge1$ be integers. On every simple, unweighted, undirected graph
with $n$ vertices, $m$ edges, and at most $k$ explicit demand pairs of
distinct vertices, a quantum adjacency-array algorithm returns the exact
minimum multicut value and an optimal vertex partition using
\[
 \OO\!\left(n+\min\{m,n\sqrt{k(n-1)}\}\right)
 =\OO\!\left(n+\min\{m,n^{3/2}\sqrt{k}\}\right)
\]
input queries, including degree queries, with success probability at
least $2/3$. The implicit constants are absolute, independent of $k$;
only polylogarithmic factors are suppressed. Neighbor arrays may have
arbitrary fixed orders. Offline computation is unrestricted and finite.
\end{theorem}

The theorem counts input queries; it does not assert a matching time
or gate bound. Even for one demand, the implementation below may
enumerate exponentially many labelings. All original-graph access is
charged, including the implementation of layer records and acquisition
of the exact quotient. Computation on explicit acquired data is free.

The proof has two main steps. Lemma~\ref{up:candidates} obtains a family
that contains every optimum and whose members are locally stable at
each vertex's original degree scale. A global cost approximation alone
does not supply this guarantee. Lemma~\ref{up:acquisition} acquires every
edge crossing the family's common refinement. It uses unions of classes
to obtain a dominant label in each large part, so exceptional vertices
charge one selected boundary without averaging over labels. Exact
optimization of the explicit quotient then completes the algorithm.

The sparsify--contract--learn framework, balanced refinements and
boundary charging build on RSW~\cite[Lemma 2.6 and Section 2.3]{RSW18}
and Apers--Lee~\cite[Lemma 2 and Section 3]{AL21}. The two lemmas
provide the degree-sensitive certificates and the uniform multicut
acquisition argument needed here. Appendix~\ref{rec:section} implements
the layer approximations on padded records using existing spectral
sampling tools. Access to a degree layer never assumes a filtered
adjacency oracle.

\subsection{Feasible labelings and degree-layer record oracles}

Let $\mathcal R$ be the demand set and $T$ its set of endpoints.
Empty demands admit the one-part partition and value zero without input
queries. Otherwise set
\[
 t=|T|,\qquad r=t\le2k,\qquad K=k(n-1)>0.
\]
A feasible labeling is $f:V\to[r]$ with $f(u)\ne f(v)$ for every demand
$\{u,v\}\in\mathcal R$. Labels may be unused. Write
\[
 \cost_G(f)=|\{uv\in E:f(u)\ne f(v)\}|,\qquad
 C=\min_{f\text{ feasible}}\cost_G(f).
\]
This optimum equals the edge-multicut optimum. In the component
partition after a feasible deletion set, at most $r$ parts contain
demand endpoints. Merge every endpoint-free part into one of those
parts. No demand becomes violated, and the number of crossing edges
cannot increase. Conversely, the crossing edges of any feasible
labeling form a feasible deletion set. Isolating one chosen endpoint
of each demand, with chosen vertices in distinct singleton parts, gives
\begin{equation}\label{up:eq:K}
 C\le k(n-1)=K,\qquad d(v)\le n-1\le K.
\end{equation}

Read all ordinary degrees using $n$ queries. For each dyadic value
$h=2^j$ with a nonempty bucket $\{v:h\le d(v)<2h\}$, concatenate
the ordinary adjacency slots of its vertices. This is a known padded
record domain of length
\begin{equation}\label{up:eq:domain}
 N_h=\sum_{v:h\le d(v)<2h}d(v)\le2nh.
\end{equation}
Order vertices lexicographically by $(d(v),v)$. A slot at $v$ with
neighbor $w$ represents the record $e_v-e_w$ when $v$ precedes $w$,
and represents the zero record otherwise. The known degree table
decides ownership after the neighbor is queried. A coherent record
query uses a constant number of original queries, including
uncomputation. Prefix sums and lookups in known tables are free
reversible computations in the query model.

Each edge has exactly one nonzero record. Its layer is determined by
\[
 h\le\min\{d(v),d(w)\}<2h.
\]
In particular, both endpoints of a layer-$h$ edge have original degree
at least $h$. Denote this layer graph by $G_h$. There are at most
$J=1+\lceil\log_2 n\rceil$ domains, their lengths sum to $2m$, and
every used $h$ satisfies $h\le n-1\le K$.
An empty record domain can be skipped. A nonempty domain can consist
entirely of zero records; Lemma~\ref{rec:main} handles this case without
requiring the algorithm to recognize it in advance. No filtered
adjacency array, inverse-neighbor search, or layer-degree oracle is used.

\subsection{Locally stable candidates from layer approximations}

For a nonnegative weighted graph $Q$, a nonterminal $v$, and a label
$b\ne f(v)$, define
\[
 \operatorname{gain}_Q(f,v,b)
 =\cost_Q(f)-\cost_Q(f^{v\to b}),
\]
where $f^{v\to b}$ changes only the label of $v$. Such a change is
always feasible when $f$ is feasible, since $v\notin T$.

\begin{lemma}\label{up:candidates}
Put $a=1/100$. For every nonempty degree-bucket domain, choose a dyadic
rational $\varepsilon_h$ satisfying
\[
 \frac a2\sqrt{\frac hK}\le\varepsilon_h
 \le a\sqrt{\frac hK},
\]
and use Lemma~\ref{rec:main} to construct a spectral approximation
$H_h$ to $G_h$. Let $H$ be the union of the $H_h$, with weights added.
Conditioned on all spectral guarantees, an explicitly computable
nonempty family $\mathcal F$ contains every optimal feasible
$r$-labeling of $G$ and satisfies
\begin{equation}\label{up:eq:candidate-properties}
 \cost_G(f)\le2K,\qquad
 \operatorname{gain}_G(f,v,b)<\frac{d(v)}4
\end{equation}
for all $f\in\mathcal F$, all nonterminals of positive degree, and
all different labels $b$. The layer approximations use
$\OO(n\sqrt K)$ original queries with a fixed worst-case budget
and total failure probability at most $1/12$.
\end{lemma}

\begin{proof}
Allocate failure probability $1/(12J)$ to each use of
Lemma~\ref{rec:main}. Since $\varepsilon_h\le a\le1/2$, this lies
in its parameter range. By~\eqref{up:eq:domain}, the query cost of
one layer, including the inverse-polylogarithmic failure parameter, is
\[
 \OO\!\left(\frac{\sqrt{N_hn}}{\varepsilon_h}\right)
 \le\OO\!\left(\frac{\sqrt{2n^2h}}{(a/2)\sqrt{h/K}}\right)
 =\OO(n\sqrt K).
\]
There are only $J$ layers; their costs sum to the same bound with
polylogarithmic factors suppressed. Exact reading within the primitive
is also permitted. All budgets are fixed even on failed outcomes.
The dyadic parameters can be chosen using exact comparisons of squared
rational quantities; no irrational input precision is needed.

Condition on all spectral guarantees. Because $\varepsilon_h\le a$,
their sum gives
\begin{equation}\label{up:eq:global-sandwich}
 (1-a)L_G\preceq L_H\preceq(1+a)L_G.
\end{equation}
We next obtain a local error bound stronger than the global sandwich
alone. Write $d_h(v)$ and $C_h(f)$ for the degree and labeling cost
in the original unweighted layer $G_h$. For a labeling $f$, vertex $v$,
and label $b\ne f(v)$, put
\[
 z=\mathbf1_{f^{-1}(f(v))}-\mathbf1_{f^{-1}(b)}.
\]
For either weighted layer graph $Q$, direct summation over incident
edges gives
\begin{equation}\label{up:eq:gain-identity}
 \operatorname{gain}_Q(f,v,b)=e_v^{\mathsf T}L_Qz-d_Q(v).
\end{equation}
A neighbor in the old label contributes zero to
$e_v^{\mathsf T}L_Qz$, a neighbor in the new label contributes twice
its edge weight, and a neighbor in any other label contributes its
edge weight. Subtracting the degree leaves new-label neighbor weight
minus old-label neighbor weight. Moreover,
\[
 z^{\mathsf T}L_{G_h}z\le4C_h(f),
\]
since edges within a label contribute zero and a crossing edge
contributes at most four.

The spectral sandwich for a layer implies, for all real vectors $x,y$,
\begin{equation}\label{up:eq:bilinear}
 |x^{\mathsf T}(L_{H_h}-L_{G_h})y|
 \le\varepsilon_h
 \sqrt{x^{\mathsf T}L_{G_h}x}\,
 \sqrt{y^{\mathsf T}L_{G_h}y}.
\end{equation}
This also holds for singular Laplacians. The sandwich gives equal
kernels, and on their common range conjugation by $L_{G_h}^{-1/2}$
has operator norm at most $\varepsilon_h$; Cauchy--Schwarz gives
the displayed inequality. Applying it to~\eqref{up:eq:gain-identity},
and applying the sandwich to the singleton degree, yields
\[
 |\operatorname{gain}_{H_h}(f,v,b)
   -\operatorname{gain}_{G_h}(f,v,b)|
 \le\varepsilon_h\bigl(2\sqrt{d_h(v)C_h(f)}+d_h(v)\bigr).
\]
Every layer incident to $v$ has $h\le d(v)$, so on these layers
$\varepsilon_h\le a\sqrt{d(v)/K}$. Using
$\sum_h d_h(v)=d(v)$, $\sum_h C_h(f)=\cost_G(f)$, and
Cauchy--Schwarz gives the uniform bound
\begin{equation}\label{up:eq:local-error}
 \left|\operatorname{gain}_H(f,v,b)-\operatorname{gain}_G(f,v,b)\right|
 \le a\,d(v)\left(2\sqrt{\frac{\cost_G(f)}K}+1\right).
\end{equation}
The degree-error terms used only $\varepsilon_h\le a$.
For an isolated vertex both gains are zero. These conclusions hold
simultaneously for all labelings and moves; no union bound over
labelings is needed.

On the explicit weighted graph $H$, compute
\[
 C_H=\min_{f\text{ feasible}}\cost_H(f),\qquad
 \rho=\frac{1+a}{1-a}.
\]
Let $\mathcal F$ consist of exactly the feasible labelings satisfying
\begin{equation}\label{up:eq:filters}
 \cost_H(f)\le\rho C_H,\qquad
 \operatorname{gain}_H(f,v,b)\le3a\,d(v)
 \quad(v\notin T,\ b\ne f(v)).
\end{equation}
All weights, gains, and thresholds are rational. The tests and the
enumeration of all $r^n$ labelings are finite offline computations.
The identity expressing labeling cost as half the sum of its
label-shore boundaries and~\eqref{up:eq:global-sandwich} give
\[
 (1-a)C\le C_H\le(1+a)C.
\]
Every original optimal labeling $f^*$ passes the cost filter, since
$\cost_H(f^*)\le(1+a)C\le\rho C_H$. All its feasible local gains
in $G$ are nonpositive. Equation~\eqref{up:eq:local-error} and
$C\le K$ imply that its gains in $H$ are at most $3a\,d(v)$,
so it passes all local filters. Thus $\mathcal F$ is nonempty
and contains every original optimal $r$-labeling.

Conversely, any retained labeling satisfies
\[
 \cost_G(f)\le\frac{\rho C_H}{1-a}\le\rho^2 C\le2K.
\]
Combining its local filter with~\eqref{up:eq:local-error} gives,
for $d(v)>0$,
\[
 \operatorname{gain}_G(f,v,b)
 \le(4+2\sqrt2)a\,d(v)<7a\,d(v)<d(v)/4.
\]
This proves~\eqref{up:eq:candidate-properties}. No division by $C$
was used, so the argument includes zero optimum.
\end{proof}

\subsection{Union shores and layerwise acquisition}

Let $\mathcal A_T$ isolate every terminal and put all nonterminals in
one part, if any. Let $\mathcal A$ be the common refinement of
$\mathcal A_T$ and all candidate labelings. It is explicit once
$\mathcal F$ is known, without further graph queries.

\begin{lemma}\label{up:acquisition}
Given the ordinary degrees and a candidate family with the properties
of Lemma~\ref{up:candidates}, all original edges crossing
$\mathcal A$ can be acquired with bounded error in
$\OO(n\sqrt K)$ queries. The algorithm has fixed worst-case query
budgets even when the candidate-family properties fail.
\end{lemma}

\begin{proof}
An empty candidate family is rejected. First read all original arrays
of source vertices in buckets with $h<\sqrt K$. Each such vertex
has degree less than $2\sqrt K$, and the buckets are disjoint, so
this costs at most $2n\sqrt K$ queries. Every edge in these low
layers is acquired through its owning record.

For each remaining nonempty domain, with $h\ge\sqrt K$, construct
its own partition $\mathcal B_h$ offline. Start with
$\mathcal A_T$ and set $s=h/8$. A permitted shore is any union
\[
 X=\bigcup_{i\in I} f^{-1}(i),\qquad f\in\mathcal F,\quad I\subseteq[r].
\]
Whenever such a shore splits a current part $P$ into two pieces
of size at least $s$, select it and refine every current part by it.
Stop when no such selection remains. All possibilities are finite and
depend only on the explicit candidate family. The potential
$\sum_P\max\{|P|-s,0\}$ is initially at most $n$. Splitting any
part cannot increase its contribution, and a balanced witness decreases
its contribution by exactly $s$. At most $8n/h$ shores are selected.

Every selected union shore has original boundary contained in the
crossing-edge set of its witnessing labeling, hence at most $2K$
edges. Let $U_h=E_G(\mathcal B_h)$, used only in the analysis.
The initial terminal singletons contribute at most $tn$ edges. Thus
\begin{equation}\label{up:eq:Wh}
 |U_h|\le tn+\frac{16nK}{h}
 \le W_h:=\frac{18nK}{h}.
\end{equation}
The last step uses $t\le2k$ and $h\le n-1$, so
$tn\le2kn\le2nK/h$. Knowing the edges of $U_h$ is not needed
to compute $\mathcal B_h$ or $W_h$.

For every part $P\in\mathcal B_h$ and every $f\in\mathcal F$, either
\begin{equation}\label{up:eq:dominant}
 |P|<3s,\quad\text{or one label occupies more than }|P|-s
 \text{ vertices of }P.
\end{equation}
Indeed, suppose $|P|\ge3s$. If a label piece has size at least
$s$, its complement must have size less than $s$, or that single
label shore would permit another balanced split. If every piece
has size less than $s$, accumulate pieces until their union first
has size at least $s$. Its size is less than $2s$ and its complement
has size at least $s$, giving a forbidden balanced union shore.
The large piece in the second case is unique; call it the dominant
label piece.

Define $Z_h$ using only known degrees and partitions. It consists
of nonterminals $v$ of degree at least $h$ in nonsingleton parts
$P\in\mathcal B_h$ for which either $|P|<3s$, or $|P|\ge3s$ and some
$f\in\mathcal F$ places $v$ outside the dominant label piece in $P$.
Let $u_v$ count the edges of $U_h$ incident to $v$. We claim
\begin{equation}\label{up:eq:charge}
 d(v)\le4u_v\quad(v\in Z_h),\qquad
 \sum_{v\in Z_h}d(v)\le8|U_h|\le8W_h.
\end{equation}
If $|P|<3s$, simplicity gives
$u_v>d(v)-3s\ge5d(v)/8$, since $s=h/8\le d(v)/8$.
Otherwise fix a witnessing labeling and its dominant label.
The vertices outside that dominant piece number less than $s$.
The number of neighbors sharing $v$'s old label is less than
$u_v+s$, whereas the number with the dominant label is greater
than $d(v)-u_v-s$. If $u_v<d(v)/4$, moving $v$ to the dominant
label would have gain greater than
\[
 d(v)-2u_v-2s\ge d(v)/4.
\]
The move is feasible because $v$ is not a terminal. It contradicts
Lemma~\ref{up:candidates}. Thus $d(v)\le4u_v$ in both cases.
Different vertices may use different candidate witnesses, but all
charges concern the same fixed $U_h$. Counting each vertex once gives
$\sum_{v\in Z_h}d(v)\le4\sum_{v\in Z_h}u_v\le8|U_h|$.

Read the entire original adjacency arrays of $Z_h$. Their total
cost over the high layers is at most
\begin{equation}\label{up:eq:volume-sum}
 \sum_{h\ge\sqrt K}8W_h
 \le144nK\sum_{h\ge\sqrt K}\frac1h
 \le288n\sqrt K,
\end{equation}
where the sum is over dyadic values. Repeated reads fit this bound
and may also be cached. No layer-specific neighbor oracle is required.

Next enumerate the nonzero records of layer $h$ whose endpoints
lie in different $\mathcal B_h$ parts. Their number is at most
$\lceil W_h\rceil$. The record oracle and the explicit part-index
table give a coherent marking predicate with a constant number of
original queries. Capped quantum enumeration and endpoint retrieval
cost
\[
 \OO\!\left(\sqrt{N_h(W_h+1)}\right)=\OO(n\sqrt K)
\]
per layer, by~\eqref{up:eq:domain} and~\eqref{up:eq:Wh}.
If the cap is at least $N_h$, read the domain instead; this is within
the same bound. Otherwise at most $\lceil W_h\rceil\le N_h$
reported records are retrieved, and retrieval is absorbed by the
enumeration bound. Allocate failure probability $1/(12J)$ to each
enumeration; amplification and the number of layers add only
polylogarithmic factors.

These acquisitions contain every original edge crossing $\mathcal A$.
An edge in a low layer was read directly. In a high layer, an edge
crossing $\mathcal B_h$ was enumerated. If a remaining edge crosses
$\mathcal A$ but lies inside a part $P$ of $\mathcal B_h$, some
candidate labels its endpoints differently. If $|P|<3s$, both
endpoints meet the small-part condition for $Z_h$; otherwise at least
one endpoint lies outside that candidate's dominant piece. Both have
degree at least $h$, and neither is a terminal because terminal parts
are singletons. Thus an endpoint belongs to $Z_h$ and its whole
adjacency array was read.

The procedure also has controlled budgets on failed sparsifier
outcomes. The set $Z_h$ and its total original degree are known
before its arrays are queried. Reject if that total exceeds $8W_h$.
Enumeration has a fixed query budget and hard caps on reported
indices and subsequent retrievals. Verify retrieved records and their
marking predicates, rejecting detected inconsistencies and
deduplicating repeated indices. An undetected enumeration failure is
included in its error probability; we do not need to recognize every
failure. Reading a whole domain when its cap exceeds its size is
also fixed in advance. Hence neither a bad candidate family nor an
unsuccessful enumeration can exceed the query budget. On the good
event,~\eqref{up:eq:charge} prevents rejection by the volume check,
and successful enumerations acquire all required edges.
The degree-volume bound, enumeration costs, and low-layer reads
prove the lemma.
\end{proof}

\subsection{The complete algorithm and output}

\begin{proof}[Proof of Theorem~\ref{up:main}]
Remove repeated demands. If none remain, return value zero and the
one-part partition. Otherwise read all degrees and obtain $m$. If
$m=0$, return zero and any feasible labeling. If $m\le n\sqrt K$,
read the graph and optimize it offline using $O(n+m)$ queries.
This includes bounded small instances and ensures the minimum with
complete input acquisition in the theorem.

On the remaining inputs, apply Lemmas~\ref{up:candidates} and
\ref{up:acquisition}. A failed primitive or a deterministic rejection
may terminate the algorithm with any fixed feasible labeling and an
arbitrary value; this outcome is counted as failure. No additional
input read is used to repair a failed branch.

On joint success, compute the full common refinement $\mathcal A$
offline. Deduplicate all acquired original edges, retain those
crossing $\mathcal A$, and aggregate them into a weighted quotient
on the parts of $\mathcal A$. Every original optimal feasible
$r$-labeling belongs to $\mathcal F$ and is constant on every such
part, so the quotient admits an original optimum. Conversely, any
feasible quotient labeling lifts to an original feasible labeling
with exactly the same cost: internal edges do not cross the lift,
and every original inter-part edge was acquired. Demand endpoints
remain distinct singleton vertices. Exact offline optimization of
this explicit quotient returns the original optimum value and an
optimal lifted partition. The acquired edges also suffice to output
its associated optimal deletion set.

All layer sparsifiers together fail with probability at most $1/12$,
and all acquisition enumerations together fail with probability at
most $1/12$. The remaining implications are deterministic on the
joint good event, so the success probability is at least $5/6$,
and hence at least $2/3$. Every branch respects the fixed query caps.
The non-read-all branch uses $\OO(n+n\sqrt K)$ queries with absolute
constants independent of the number of labels and demands. In that
branch $K<(m/n)^2\le n^2$, so all logarithmic precision and record
parameters are bounded by polylogarithmic functions of $n$.
Combining both branches gives
\[
 \OO\!\left(n+\min\{m,n\sqrt K\}\right).
\]
For $n\ge2$, $n-1$ is within a factor two of $n$, giving the second
form in Theorem~\ref{up:main}.
\end{proof}

\begin{corollary}\label{up:three-terminal}
For three distinct prescribed terminals in a simple unweighted graph,
exact minimum three-terminal multiway cut, including its value and an
optimal three-partition, admits a quantum adjacency-array algorithm
using $\OO(n+\min\{m,n^{3/2}\})$ queries with success probability
at least $2/3$.
\end{corollary}

\begin{proof}
Use the three pairwise terminal demands, so $k=3$. There are three
available labels, and feasibility forces the terminal labels to be
distinct. Apply Theorem~\ref{up:main}.
\end{proof}

\paragraph{Uniform parameter dependence.}
The constant $a=1/100$ and the $J=O(\log n)$ layers do not depend on
the label count. The bounds $N_h\le2nh$,
$\varepsilon_h=\Theta(\sqrt{h/K})$, $W_h=18nK/h$, and
$\sum_{v\in Z_h}d(v)\le144nK/h$ account for every input-query
contribution. The number $r\le2k$ of labels affects offline enumeration
and union-shore selection, but introduces no further query factor.

The proof covers disconnected graphs, isolated vertices, zero optimum,
and singular layer Laplacians. Its use of vertex cardinalities to bound
same-label neighbor counts requires simplicity and unit original edge
weights; the weighted layer approximations are only auxiliary graphs.
All candidate enumeration, union-shore selection, rational linear
algebra, and quotient optimization are finite offline computations,
but may require exponential time. The theorem asserts an input-query
bound, not a corresponding total running-time guarantee.

\section{A common construction for the lower bounds}\label{sec:lower-tools}

The global-cut and multicut lower bounds use the same degree-preserving
switches. Their difference is how a threshold is imposed: a fixed attachment
competes with the core cut in the global problem, while selector vertices
with separation demands perform this role for multicut. We first record the
common bit-query lower bound and graph representation. Hamming-weight
counting here is a tool for finding \emph{one} optimum cut; no result about
counting the family of minimum cuts is needed.

\subsection{A Hamming-layer lower bound}

We use the positive-weight adversary theorem~\cite{Ambainis02}; related
counting bounds also follow from~\cite{NayakWu99}.

\begin{lemma}\label{lo:counting}
Let $N\ge2$ and $1\le p\le N-1$ be integers. Distinguishing Boolean
strings of Hamming weight $p-1$ from strings of Hamming weight $p+1$,
with success probability at least $2/3$ on every promised input, requires
\[
 \Omega\!\left(\sqrt{(N-p+1)(p+1)}\right)
\]
quantum bit queries.
\end{lemma}

\begin{proof}
Let $X,Y$ be the two Hamming layers. Relate $x\in X$ to $y\in Y$ when
two zero bits of $x$ are changed to one. Every $x$ has
$a=\binom{N-p+1}{2}$ related inputs, and every $y$ has
$b=\binom{p+1}{2}$. For a specified input and coordinate, the number of
related inputs differing there is at most $N-p$ on the $X$ side and
at most $p$ on the $Y$ side. The adversary bound is therefore
\[
 \Omega\!\left(\sqrt{\frac{ab}{(N-p)p}}\right)
 =\Omega\!\left(\frac12\sqrt{(N-p+1)(p+1)}\right).
\]
All factors are positive in the stated range, including $p=1$ and $p=N-1$.
\end{proof}

\subsection{The four-clique switch core}

For an integer $q\ge2$ and $x\in\{0,1\}^{q\times q}$, let $H_q(x)$
have four disjoint cliques
\[
 A=\{a_1,\ldots,a_q\},\quad B=\{b_1,\ldots,b_q\},\quad
 C=\{c_1,\ldots,c_q\},\quad D=\{d_1,\ldots,d_q\}.
\]
All edges have weight one. For each $(i,j)\in[q]^2$, add the two edges
\[
\begin{array}{c|cc}
 x_{i,j}&\text{first edge}&\text{second edge}\\\hline
 0&a_i d_j&c_i b_j\\
 1&a_i b_j&c_i d_j.
\end{array}
\]
Switches of this form preserve degrees and are related to the graph-query
constructions in~\cite{AL21,BGMP21}. We give all properties used here.

\begin{lemma}\label{sw:core}
The graph $H_q(x)$ is simple, has $4q$ vertices and $4q^2-2q$ edges,
and every vertex has degree $2q-1$, independently of $x$.
Writing $t=|x|$, the nontrivial cuts that do not split a clique have the
following costs, with complementary shores identified:
\[
\begin{array}{c|c}
 \text{shore}&\text{cost}\\\hline
 A,B,C,D&q^2\\
 A\cup C&2q^2\\
 A\cup B&2(q^2-t)\\
 A\cup D&2t.
\end{array}
\]
A cut splitting a clique has cost at least $q-1$.
If $0<t<q^2$, the graph is connected.
\end{lemma}

\begin{proof}
Each switch uses a different pair of indices, and its edges join different
cliques. Hence there are no repeated edges or overlaps with clique edges.
Every vertex has $q-1$ clique neighbors and $q$ switch neighbors. The edge
count is $4\binom q2+2q^2$. Each switch contributes one edge incident to
each clique, two across $(A\cup C,B\cup D)$, and either two or zero across
each of the other two bipartitions, as recorded in the table. These are all
seven unordered nontrivial cuts of four groups. Splitting a clique into
parts of sizes $s,q-s$ contributes $s(q-s)\ge q-1$ edges. Finally, a zero
switch joins $A$ to $D$ and $B$ to $C$, and a one switch joins those two
connected pairs. The cliques themselves are connected.
\end{proof}

\begin{lemma}\label{sw:oracle}
There is a fixed adjacency-array representation of $H_q(x)$ for which each
degree or neighbor query can be simulated coherently with at most two bit
queries to $x$. The same holds after adjoining known vertices and non-switch edges,
or making known changes to clique edges, provided the added or removed
neighbors occupy input-independent positions relative to the switch slots.
\end{lemma}

\begin{proof}
List the clique neighbors in a fixed order, followed by the $q$ switch
neighbors. The switch slots are
\[
\begin{array}{c|cc}
 \text{vertex}&x_{i,j}=0&x_{i,j}=1\\\hline
 a_i&d_j&b_j\\
 b_j&c_i&a_i\\
 c_i&b_j&d_j\\
 d_j&a_i&c_i.
\end{array}
\]
The slot index is $j$ at $a_i,c_i$ and $i$ at $b_j,d_j$. A queried vertex
and switch slot therefore determine one bit coordinate. Query that bit into
an ancilla, compute the neighbor label reversibly, and query the bit again
to erase the ancilla. Uncompute all temporary indices. Known clique or
auxiliary slots, degrees, and invalid-slot answers are independent of $x$.
On such branches use any fixed valid bit index, ignore its value when
forming the answer, and erase it with the second query. Thus the simulation
also works on superpositions of vertices, slots, and query types without
assuming controlled access to an unspecified oracle. Known changes to
non-switch slots only change fixed offsets.
\end{proof}

The next section uses all $q^2$ bits and a single fixed attachment. The
multicut construction in Section~\ref{sec:lower} restricts the variable
bits to a smaller block and attaches several demand selectors. This keeps
the query simulation unchanged while altering the optimal-partition
threshold.

\section{The global minimum-cut lower bound}\label{sec:global-lower}

A global cut chooses its two shores freely; it is not the one-demand
special case of multicut. We show that even without prescribed terminals,
finding one optimum already requires $\Omega(n^{3/2})$ adjacency-array
queries on dense unweighted graphs.

\begin{theorem}\label{glob:main}
For every integer $n\ge33$, there is a connected simple unweighted graph
family on exactly $n$ vertices and $\Theta(n^2)$ edges on which either
computing the global minimum-cut value or outputting any minimum-cut shore
requires $\Omega(n^{3/2})$ bounded-error quantum adjacency-array queries.
The labeled degree sequence and edge count are input-independent and may
be supplied for free. The lower bound holds even for sorted neighbor arrays.
\end{theorem}

Allowing $O(n)$ degree preprocessing and full reading when $m<n$, the
existing global-cut algorithm of Apers--Lee~\cite[Theorem 21]{AL21} gives
$\OO(n+\sqrt{mn})$ queries. Together with our lower bound this gives
dense worst-case complexity
$\widetilde\Theta(n^{3/2})$ for both tasks. The upper bound in this
comparison is not a new algorithm of this paper. Theorem~\ref{glob:main}
rules out the near-linear query complexity proposed for simple unweighted
adjacency-array inputs in~\cite[Section 1.3]{AL21}. It does not characterize
the lower bound for each sparse edge count.

\subsection{A fixed threshold for the global optimum}

Fix $q\ge8$ and set
\[
 N=q^2,\qquad r=\lfloor q/2\rfloor,\qquad
 p=\lfloor r/2\rfloor=\lfloor q/4\rfloor.
\]
Promise that $x\in\{0,1\}^{q\times q}$ has weight $p-1$ or $p+1$.
Start with $H_q(x)$ from Section~\ref{sec:lower-tools}, add a vertex $z$,
and join $z$ to $a_1,\ldots,a_r$. Call the resulting graph $G_x$.
It has
\[
 n_0=4q+1,\qquad m_0=4q^2-2q+r.
\]
The selected $A$-vertices have degree $2q$, all other core vertices have
degree $2q-1$, and $z$ has degree $r$. These quantities do not depend on
the input. Both promised layers satisfy $0<|x|<q^2$, so the graph is
connected by Lemma~\ref{sw:core}.

\begin{lemma}\label{glob:classification}
The minimum cut of $G_x$ is unique up to complement. Its value and a shore
are given by
\[
\begin{array}{c|c|c}
 |x|&\lambda(G_x)&\text{minimum-cut shore}\\\hline
 p-1&2(p-1)&A\cup D\cup\{z\}\\
 p+1&r&\{z\}.
\end{array}
\]
\end{lemma}

\begin{proof}
The cut isolating $z$ has size $r$. Any cut splitting one of the four
cliques has size at least $q-1>r$. Thus an optimum puts each clique on
one side in its entirety. If both sides contain core cliques, placing
$z$ with $A$ leaves the core cut cost unchanged, whereas placing it
opposite $A$ adds $r>0$. If all four cliques are on the same side, the
only nontrivial cut isolates $z$.

For $t=|x|$, the remaining core cut costs are listed in
Lemma~\ref{sw:core}. The entries $N$ and $2N$ exceed $r$, as does
\[
 2(N-t)\ge 2N-2(p+1)\ge 2N-r-2>r.
\]
The last inequality holds since $N=q^2$ and $r\le q/2$. Only the cut
with shore $A\cup D\cup\{z\}$, of size $2t$, and the cut isolating
$z$, of size $r$, can therefore minimize. Finally,
\[
 2(p-1)<r<2(p+1).
\]
These inequalities are strict whether $r$ is odd or even, and establish
both the values and uniqueness.
\end{proof}

\subsection{Query simulation, padding, and output discrimination}

The representation in Lemma~\ref{sw:oracle} extends to $G_x$ by appending
$z$ to the arrays of $a_1,\ldots,a_r$, and giving $z$ the fixed array
$a_1,\ldots,a_r$. Thus every graph query uses at most two bit queries.

The lower bound also permits sorted arrays. Give the original vertices
the label order
\[
 a_1,c_1,a_2,c_2,\ldots,a_q,c_q,
 b_1,d_1,b_2,d_2,\ldots,b_q,d_q,
\]
and place $z$ afterwards. At an $a_i$ or $c_i$, fixed clique neighbors
precede exactly one neighbor from each consecutive pair $\{b_j,d_j\}$.
At a $b_j$ or $d_j$, exactly one member of each $\{a_i,c_i\}$ precedes
the fixed clique neighbors. Therefore the position of a switch neighbor
does not depend on other bits. A sorted switch slot still determines a
single bit, and the clean simulation in Lemma~\ref{sw:oracle} applies.

\begin{proof}[Proof of Theorem~\ref{glob:main}]
First consider $n=4q+1$. A correct value output distinguishes the two
input layers by testing whether it is less than $r$. A correct shore
output distinguishes them by testing whether it or its complement is
$\{z\}$. Lemma~\ref{glob:classification} makes this test valid for
\emph{every} correct output, with no query to evaluate its cost.
The clean simulation and Lemma~\ref{lo:counting} imply the lower bound
\[
 \Omega\!\left(\sqrt{(N-p+1)(p+1)}\right)=\Omega(q^{3/2}),
\]
since $N=q^2$, $p\ge q/8$, and $N-p+1\ge N/2$ for $q\ge8$.

For an arbitrary $n\ge33$, set
\[
 q=\left\lfloor\frac{n-1}{4}\right\rfloor,\qquad
 s=n-(4q+1)\in\{0,1,2,3\}.
\]
Adjoin $s$ vertices to $A$ and complete the enlarged clique, while only
its original $q$ vertices participate in switches or attach to $z$.
A cut splitting this enlarged group has at least $q+s-1>r$ clique
edges; all whole-group cut costs are unchanged. The classification
therefore remains valid, using the enlarged $A$ in the first shore.
The new edges, degrees, and slot offsets are fixed independently of $x$.
For sorted arrays put the new labels after all original core labels and
before $z$; their known clique slots do not alter the local switch rule.
The edge count becomes
\[
 4q^2-2q+r+sq+\binom{s}{2}=\Theta(n^2).
\]
The graph remains connected, $q=\Theta(n)$ with absolute constants, and
the same output tests apply. This proves all claims.
\end{proof}

\subsection{Regular instances with balanced optimum cuts}

The threshold vertex is a convenient proof device, but the lower bound
does not depend on a low-degree singleton being optimal. The following
strengthening concerns the same single-output global-cut task.

\begin{proposition}\label{glob:regular}
For every $q\ge8$ divisible by eight, there is a connected simple
unweighted family on $n=6q$ vertices, all of degree $2q-1$, for which
computing the global minimum-cut value or any optimum shore requires
$\Omega(n^{3/2})$ quantum queries. Each input has a unique minimum cut
up to complement, with shore sizes $2q$ and $4q$. The lower bound holds
with sorted neighbor arrays and free knowledge of the degree sequence.
\end{proposition}

\begin{proof}
Start with $H_q(x)$ and add a clique $Z$ of size $2q$. Set $r=q/2$ and
$p=q/4$. Choose $r$ vertices in each of $A$ and $Z$, delete a fixed
perfect matching on each selected set, and join the selected sets by a
fixed matching of $r$ edges. The selected sets have even size. Every
affected vertex loses one edge and gains one, and all vertices have
degree $2q-1$.

A cut splitting $A$ has at least $q-2$ surviving clique edges: for a
subset of size $u$, deleting a matching removes at most
$\min\{u,q-u\}$ of its $u(q-u)$ boundary edges. The corresponding
bound for $Z$ is $2q-2$; for each other clique it is $q-1$. All these
bounds exceed $r$, the cost of isolating $Z$. Thus no optimal cut
splits any group. For cuts of entire groups the table of
Lemma~\ref{sw:core} is unchanged, with a cost of $r$ added precisely
when $Z$ is opposite $A$. Under $|x|=p\pm1$, the only candidates are
$A\cup D\cup Z$ of cost $2|x|$ and $Z$ of cost $r$. Since
$2(p-1)=r-2<r<r+2=2(p+1)$, the optimum is unique, with the stated
shore sizes. Both switch types occur, all internal groups remain
connected, and the fixed matching joins $Z$ to $A$.

An optimum shore distinguishes the layers by whether it or its complement
equals $Z$; its value distinguishes them by comparison with $r$.
Use the same interleaved core labels as above, putting $Z$ last.
All deleted and added edges are fixed, so only known offsets change;
each sorted switch slot still depends on one bit. Lemmas~\ref{sw:oracle}
and~\ref{lo:counting} give $\Omega(q^{3/2})=\Omega(n^{3/2})$ queries.
\end{proof}

This regularity statement is on the grid $n=6q$, $8\mid q$.
Theorem~\ref{glob:main} supplies the separate guarantee for every
$n\ge33$. Neither regularity nor sorted-array hardness is assumed
for the multicut family in Section~\ref{sec:lower}.

\section{A lower bound with a growing number of demands}\label{sec:lower}

We extend the threshold construction of Section~\ref{sec:global-lower}
to prescribed demands, using the same core $H_q(x)$ and bit-query lower
bound from Section~\ref{sec:lower-tools}. The global proof need only
classify bipartitions. Here an optimum may use more than two parts, so
we first restrict the variable switches to make two sides nearly complete,
then attach selectors whose demands impose the desired threshold. The
classification below covers all feasible partitions and is the additional
step needed for an output-sensitive multicut lower bound.

\begin{theorem}\label{lo:main}
For every $n\ge5120$ and integer $1\le k\le\lfloor n/2\rfloor$, there is a
family of connected simple unweighted undirected graphs on exactly $n$
vertices, with at most $k$ pairwise vertex-disjoint demands and
$m=\Theta(n^2)$, requiring $\Omega(n^{3/2}\sqrt{k})$ quantum adjacency-array
queries for any one of the following outputs: the exact optimum value, an
optimal partition, or a minimum-cardinality feasible edge-removal set.
The implicit constants are absolute. Within each family the degrees, edge
count, and demands are input-independent and may be given for free.
\end{theorem}

We first prove the explicit family below. Its actual demand count is denoted
by $\ell$, reserving $k$ for the demand budget of Theorem~\ref{lo:main}.
The final subsection supplies the padding needed for all values
of $n$ and $k$ in the theorem.

\begin{proposition}\label{lo:family}
Let $q\ge 1024$ be divisible by $256$, and let $1\le \ell\le q/2$ be an integer.
There is a family of connected, simple, unweighted, undirected Multicut
instances with $\ell$ pairwise vertex-disjoint demand pairs and
\[
 n=4q+\ell,\qquad m=4q^2-2q+\frac{\ell q}{128},
\]
such that every bounded-error quantum algorithm requires
$\Omega(q^{3/2}\sqrt{\ell})=\Omega(n^{3/2}\sqrt{\ell})$ adjacency-array queries
to output any one of the following: the exact optimum value, an optimal
partition, or a minimum-cardinality feasible edge-removal set.
The labeled degree sequence, the demands, and the number of edges are the
same throughout the family and may be supplied for free.
\end{proposition}

This proposition describes an explicit parameter grid, rather than every feasible
triple $(n,m,\ell)$.  In this grid $m=\Theta(n^2)$.  Taking $\ell=q/2$ gives
$\ell=\Theta(n)$ and the lower bound $\Omega(n^2)=\Omega(m)$.
The adjacency arrays used in the reduction have fixed slot layouts; they
are not asserted to be sorted by neighbor identifier.

\subsection{The graph and its query representation}

Set
\[
 r=\frac q8,\qquad L=\frac q{128},\qquad
 N=r^2,\qquad p=\frac{\ell L}{2}-1.
\]
In this subsection, $N$ is the number of variable bits.
Let $y\in\{0,1\}^{r\times r}$ have weight $p-1$ or $p+1$.
Extend it to a $q\times q$ matrix $x$ by placing $y$ in
$[r]\times[r]$, setting $x_{q,q}=1$, and setting all remaining entries
to zero.  The fixed one lies outside the variable submatrix.

Take the core $H_q(x)$ from Lemma~\ref{sw:core}, with groups $A,B,C,D$.
Every core vertex has degree $2q-1$. Define
\[
 S=A\cup D,\qquad T=B\cup C.
\]
The induced graphs on $S$ and $T$ are copies of $K_{2q}$ with edges
deleted.  The maximum deleted degree is at most
$\delta=r=q/8$, because a row or column has at most $r$ variable ones,
while the fixed one occupies its own row and column.  The maximum degree
in the bipartite graph of edges between $S$ and $T$ is also at most
$\delta$.  Its total number of edges is
\begin{equation}\label{lo:E}
 E=2(1+|y|)=\ell L-2\quad\text{or}\quad \ell L+2.
\end{equation}

Add new vertices $Z=\{z_1,\ldots,z_\ell\}$.  For every $i\in[\ell]$,
join $z_i$ to every vertex of
$U=\{a_1,\ldots,a_L\}$, and introduce the demand pair $(z_i,c_i)$.
There are no other edges incident to $Z$.  In particular, the demand
pairs are vertex-disjoint, even though all selectors have the same
neighborhood.

The graph is simple: the four clique edge sets and the switch edge sets
are disjoint, each switch uses its own pair of indices, and selector
edges have a new endpoint.  It is connected.  Indeed, $A$ and $D$ are
connected to one another by at least one zero switch, as are $B$ and $C$;
the fixed one joins these two connected subgraphs; and each selector
has $L\ge1$ neighbors.  Counting clique, switch, and selector edges gives
\[
 m=4\binom q2+2q^2+\ell L=4q^2-2q+\ell L.
\]
The degree of $a_i$ is $2q-1+\ell$ for $i\le L$ and $2q-1$ otherwise.
Every other core vertex has degree $2q-1$, and every selector has degree
$L$.  Thus no degree depends on $y$.

Use the core arrays from Lemma~\ref{sw:oracle}. For $a_i\in U$,
append $z_1,\ldots,z_\ell$ in that order; each $z_i$ has the fixed array
$a_1,\ldots,a_L$. A switch slot determines a coordinate of $x$ locally.
Inside $[r]^2$ it queries the corresponding bit of $y$; outside it the
answer is fixed. Compute--answer--uncompute uses two queries to $y$.
On fixed-answer branches use any fixed variable coordinate, ignore its
value, and uncompute it, as in Lemma~\ref{sw:oracle}. Thus degree queries,
invalid slots, and arbitrary superpositions of vertices, slots, and query
types all admit a clean two-query simulation. This is an arbitrary fixed
array representation; we do not need a sorted-array guarantee here.

\begin{figure}[t]
\centering
\begin{tikzpicture}[font=\small,>=Latex,cluster/.style={draw,rounded corners,minimum width=2.8cm,minimum height=2.5cm,fill=blue!3},dot/.style={circle,draw,fill=white,inner sep=2pt}]
\node[cluster] (S) at (0,0) {};
\node[cluster] (T) at (6.5,0) {};
\node at (0,0.68) {$S=A\cup D$};
\node at (6.5,0.68) {$T=B\cup C$};
\node[draw,rounded corners,fill=blue!12,inner sep=5pt] (U) at (0,-0.65) {$U\subseteq A$, $|U|=L$};
\node[dot] (c) at (6.5,-0.65) {$c_i$};
\node[dot] (z) at (3.25,-2.0) {$z_i$};
\draw[line width=1pt] (U.south east) -- node[below left,pos=.48] {$L$ graph edges} (z.north west);
\draw[dashed,red!65!black,line width=.8pt] (z.north east) -- node[below right,pos=.53] {demand $(z_i,c_i)$} (c.south west);
\draw[<->,line width=1pt] (S.east) -- node[above] {$E=2(1+|y|)$} node[below] {core crossing edges} (T.west);
\node[align=center] at (3.25,-3.0) {$\ell$ selector vertices share $U$; the $c_i$ are distinct.\\Two candidate costs: $E$ and $\ell L$.};
\end{tikzpicture}
\caption{Schematic of the lower-bound construction. The two core clusters
are almost cliques. Solid connections represent graph edges; the dashed
connection is a separation demand, not a graph edge. The picture shows one
of the $\ell$ selector vertices. The proof establishes that no mixed partition
beats either candidate solution.}
\label{fig:selector}
\end{figure}

\subsection{Classification of all optimal solutions}

For a vertex partition $\Pi$, its cost is the number of edges with
endpoints in distinct parts.  We consider arbitrary numbers of parts.
Call a part a strict majority part of $S$, or of $T$, if it contains
more than half the vertices of that set.

\begin{lemma}\label{lo:classification}
For the constructed graph and demands,
\[
 \operatorname{OPT}=\min\{E,\ell L\}.
\]
If $|y|=p-1$, every optimal partition has precisely the two nonempty
parts $S\cup Z$ and $T$.  If $|y|=p+1$, every optimal partition places
all core vertices in one part and every selector outside that part;
the selectors may be grouped arbitrarily among themselves.
\end{lemma}

\begin{proof}
Two feasible partitions give upper bounds.  The partition
$\{S\cup Z,T\}$ has cost $E$.  The partition consisting of $S\cup T$
and the selector singletons has cost $\ell L$.

First suppose that some cluster $W\in\{S,T\}$ has no strict majority
part.  Write $u_1,u_2,\ldots$ for the sizes of its intersections with
the parts of $\Pi$.  Since $|W|=2q$ and each $u_j\le q$, the number of
crossing edges in a complete graph on $W$ is
\[
 \frac{(2q)^2-\sum_j u_j^2}{2}\ge q^2.
\]
At most $(2q)\delta/2=q\delta$ edges were deleted from that complete
graph.  Hence $\Pi$ cuts at least
\[
 q^2-q\delta=\frac{7q^2}{8}>\ell L,
\]
where $\ell L\le q^2/256$.  Such a partition cannot be optimal.

We may therefore assume that both clusters have strict majority parts.
Let $P_S,P_T$ denote those parts, and set
\[
 r_S=|S\setminus P_S|,\qquad r_T=|T\setminus P_T|,
 \qquad R=r_S+r_T.
\]
The core edges inside $S$ crossing from $S\cap P_S$ to its complement
number at least $r_S(2q-r_S-\delta)\ge(q-\delta)r_S$.
The analogous bound holds for $T$.  Thus the internal core contribution
to the partition cost is at least $(q-\delta)R$.

Suppose first that $P_S\ne P_T$.  An edge between $S$ and $T$ can be
uncut only if at least one endpoint is outside its own cluster's
majority part.  Since the cross degree is at most $\delta$, at most
$\delta R$ of the $E$ cross edges can be saved.  Therefore
\begin{equation}\label{lo:distinct}
 \operatorname{cost}(\Pi)\ge E+(q-2\delta)R.
\end{equation}
If $R>0$, this is strictly greater than $E$ and cannot improve the
candidate of cost $E$.  If $R=0$, the core is partitioned as $S,T$.
To attain cost $E$, all selector edges must remain uncut; since $L>0$
and all their neighbors lie in $S$, every selector must belong to the
part containing $S$.

Suppose instead that $P_S=P_T=P$.  We bound the number of selector
edges that remain uncut.  When $z_i\notin P$, any uncut edge from
$z_i$ must end at a vertex of $U\setminus P$, of which there are at
most $r_S$.  When $z_i\in P$, feasibility forces $c_i\notin P$.
The vertices $c_i$ are distinct, so there are at most $r_T$ selectors
in $P$, each with only $L$ incident edges.  Altogether, at most
$\ell r_S+Lr_T$ selector edges are uncut.  Ignoring the nonnegative cost
of edges between $S$ and $T$ gives
\begin{equation}\label{lo:same}
 \begin{split}
 \operatorname{cost}(\Pi)
 &\ge (q-\delta)(r_S+r_T)+\ell L-\ell r_S-Lr_T\\
 &=\ell L+(q-\delta-\ell)r_S+(q-\delta-L)r_T.
 \end{split}
\end{equation}
Both coefficients are positive: $q-\delta-\ell\ge3q/8$, and
$q-\delta-L>0$.  Consequently $R>0$ gives cost strictly greater
than $\ell L$.  If $R=0$, all core vertices lie in $P$, so feasibility
puts every selector outside $P$ and the cost is exactly $\ell L$.

The preceding cases prove that the minimum is $\min\{E,\ell L\}$.
For the low Hamming layer, $E=\ell L-2$, so only the distinct-majority
case with $R=0$ can be optimal, and all selectors must join $S$.
For the high layer, $E=\ell L+2$, so only the common-majority case with
$R=0$ can be optimal.  This also proves the claimed classification.
\end{proof}

\subsection{Completing the reduction}

\begin{proof}[Proof of Proposition~\ref{lo:family}]
The counting parameters satisfy
\[
 p+1=\frac{q\ell}{256}\ge4,\qquad
 p\le\frac{q^2}{512}-1=\frac N8-1.
\]
Thus both Hamming layers are nonempty and Lemma~\ref{lo:counting}
applies, giving
\[
 \Omega\!\left(\sqrt{(N-p+1)(p+1)}\right)
 =\Omega\!\left(q^{3/2}\sqrt{\ell}\right)
\]
bit queries.  The graph has the claimed parameters and structural
properties, and every graph query costs at most two bit queries.

An exact optimum value distinguishes the layers because the two
values are $\ell L-2$ and $\ell L$.  An optimal partition distinguishes them
without an additional graph query: by Lemma~\ref{lo:classification},
$a_1$ and $c_1$ belong to different parts in the low layer and to the
same part in the high layer.

For completeness, suppose the output is a minimum-cardinality
feasible edge-removal set $F$.  Let $\Pi_F$ be the partition into
connected components of the graph after removing $F$.  Every edge
crossing $\Pi_F$ lies in $F$.  Conversely, if an edge of $F$ has both
endpoints in the same component, restoring it leaves all component
memberships unchanged and preserves feasibility, contradicting the
minimum cardinality of $F$.  Thus $F$ is exactly the set of crossing
edges of $\Pi_F$, and $\Pi_F$ is an optimal partition.  The known
selector edge $z_1a_1$ is absent from every optimal removal set in the
low layer and present in every such set in the high layer.  Its
membership in the classical output therefore distinguishes the
layers, without querying the graph or computing components.

In each output convention, a successful graph algorithm gives a
successful algorithm for the promised counting problem.  The bit-query
lower bound and the constant-query simulation imply the claimed graph
query lower bound.  Finally, $4q+1\le n\le(9/2)q$, so replacing $q$
by $n$ changes only an absolute constant.
\end{proof}

\subsection{Uniform vertex counts and demand budgets}
\begin{proof}[Proof of Theorem~\ref{lo:main}]
For prescribed $n\ge5120$ and $1\le k\le\lfloor n/2\rfloor$, choose
\[
 q=256\left\lfloor\frac n{1280}\right\rfloor,
 \qquad \ell=\min\{k,q/2\}.
\]
Then $q\ge1024$ is divisible by $256$, $1\le\ell\le q/2$, and
$3n/20\le q\le n/5$. The lower inequality follows from
$q>n/5-256$ and $256\le n/20$.
The family in Proposition~\ref{lo:family} has
$n_0=4q+\ell\le(9/2)q\le9n/10$ vertices.
Attach $p_0=n-n_0$ new leaves to $a_1$, with no new demands.
The graph now has exactly $n$ vertices, remains simple and connected, and has
\[
 m=4q^2-2q+\ell q/128+p_0=\Theta(n^2).
\]
The leaves have degree one, $a_1$ gains the known degree $p_0$, and all
degrees are still independent of the hidden bits.

Every leaf shares the part of $a_1$ in every optimal partition: moving a
leaf into that part removes its sole crossing edge and violates no demand.
Conversely, any core partition extends with no extra cost by placing all
leaves with $a_1$. Thus padded and unpadded optimum values agree, and the
restriction of every optimal padded partition is optimal in the original
instance. An optimal removal set contains no leaf edge, since restoring
one cannot reconnect a demanded pair and would reduce the set's size.
All three output discriminators from Proposition~\ref{lo:family} persist.

Append the new leaves after the existing entries in the array of $a_1$;
each leaf's sole entry is $a_1$. These entries and all degree changes are
known independently of the input bits. Existing switch positions are
unchanged, so the same two-bit-query simulation works coherently, also
for leaf vertices and invalid slots.

The resulting lower bound is $\Omega(q^{3/2}\sqrt\ell)$. If $k\le q/2$,
then $\ell=k$ and $q\ge3n/20$ gives the result. Otherwise $\ell=q/2$,
and $k\le n/2$ yields
\[
 q^{3/2}\sqrt\ell=\frac{q^2}{\sqrt2}
 \ge\frac{9n^2}{400\sqrt2}
 \ge\frac9{400}n^{3/2}\sqrt k.
\]
The constants are therefore uniform in both parameters.
\end{proof}

For fixed positive $k$, this is a dense $\Omega(n^{3/2})$ lower bound,
including partition-only output. For linear demand budgets in the stated
range, it becomes $\Omega(n^2)=\Omega(m)$ and matches full acquisition.
The two optimum values still differ by only two; the result does not imply
a fixed-factor approximation lower bound.

\section{Consequences and further questions}\label{sec:discussion}
The global-cut lower bound and the multicut theorem describe the cost of
finding one optimum under different separation requirements. Global cut
and prescribed $s$--$t$ cut both have dense worst-case complexity
$\widetilde\Theta(n^{3/2})$, by separate results. For multicut, a growing
demand budget increases this to $\widetilde\Theta(n^{3/2}\sqrt{k})$
throughout $1\le k\le\lfloor n/2\rfloor$. A linear demand budget
requires full acquisition in order.

The lower bounds use one graph-query mechanism: degree-preserving switches
hide a Hamming weight while a competing cut fixes a threshold. The global
construction forces a unique optimum to change, and the multicut selectors
force every optimum partition to reveal the threshold. Sharing this
representation does not identify the two optimization problems; their
different feasible sets require separate classification arguments.

The upper proof separates two requirements for exact recovery. A candidate
family must retain an original optimum, and the edges crossing its common
refinement must be inexpensive to acquire. Degree-sensitive estimates of
local improvement address the first requirement while imposing useful
stability on every candidate. Refinement by unions of classes turns that
stability into a uniform acquisition bound. This identifies the structure
needed for extensions: the local moves must preserve feasibility, and
exceptional vertices must charge a common boundary without a loss in the
number of labels.

Several questions remain outside these results. The dense lower bounds do
not characterize the dependence on a prescribed sparse edge count $m$.
Extending the multicut upper bound to weighted inputs requires additional
structure because a small vertex set
can contain a large amount of incident weight, invalidating the cardinality
bound used in the local-improvement argument. Efficient total-time
algorithms would need to replace exhaustive candidate enumeration and
common refinement, as well as exact offline quotient optimization. Finally,
these lower bounds distinguish nearby exact values and do not establish
fixed-factor approximation hardness. Each of these questions requires
an ingredient beyond the dense query bounds proved here.

\appendix
\section{Spectral sparsification from padded edge records}\label{rec:section}

We require a sparsifier for an oracle that reveals edge records, rather
than the neighbors of a specified vertex in a filtered graph. We give
the implementation and its query analysis explicitly. The recursive
sampling mechanism is related to repeated halving~\cite{Cohen14}; the
following presentation includes the estimates and failure-path bounds
used here. We include a self-contained implementation for padded records,
including singular Laplacians and fixed query budgets on failure branches.

A record is either zero or an ordered pair of distinct vertices of $[n]$,
representing the incidence vector $b_i=e_u-e_v$. Repeated and zero records
are allowed. One oracle query reveals a record coherently. For an index
set $I\subseteq[N]$, put
\[
 L_I=\sum_{i\in I}b_i b_i^{\mathsf T}.
\]
All linear algebra, storage and reversible lookup of acquired data, and
generation and storage of input-independent random tables are free.
The construction never requests a list of records incident to a given
vertex.

\begin{lemma}[Edge-record spectral sparsification]\label{rec:main}
Let $n\ge2$, $N\ge1$, $0<\varepsilon\le1/2$, and $0<\delta\le1/4$.
From a coherent oracle for $N$ incidence records on $n$ vertices, one
can obtain an explicit nonnegative rationally weighted sublist with
Laplacian $\widehat L$ such that, with probability at least $1-\delta$,
\[
 (1-\varepsilon)L_{[N]}\preceq\widehat L
 \preceq(1+\varepsilon)L_{[N]}.
\]
The algorithm has a fixed worst-case query budget
\[
 \OO\!\left(\frac{\sqrt{Nn}}{\varepsilon\sqrt\delta}\right),
\]
where suppressed factors are logarithmic in $N$, $n$, and $1/\delta$.
Reading all $N$ records is an alternative. In particular, for
$N\le n^{O(1)}$ and inverse-polylogarithmic $\delta$, the query bound is
$\OO(\min\{N,\sqrt{Nn}/\varepsilon\})$.
Both the output size and the queries used to retrieve output records
are explicitly capped. Offline computation is finite but otherwise
unrestricted.
\end{lemma}

For a positive semidefinite matrix $M$, define
\[
 R_M(b)=
 \begin{cases}
 b^{\mathsf T}M^\dagger b,& b\in\operatorname{im}M,\\
 +\infty,&b\notin\operatorname{im}M,
 \end{cases}
 \qquad
 \sigma_M(b)=\min\{1,R_M(b)\}.
\]
Here $R_M(0)=0$, including when $M=0$, and
$\min\{1,+\infty\}=1$. For every $i\in I$, $b_i\in\operatorname{im}L_I$,
$R_{L_I}(b_i)\le1$, and
\begin{equation}\label{rec:rank-identity}
 \sum_{i\in I}R_{L_I}(b_i)=\operatorname{rank}L_I\le n.
\end{equation}
Indeed, if $B$ has rows $b_i^{\mathsf T}$, these quantities are the
diagonal entries of the orthogonal projection $B(B^{\mathsf T}B)^\dagger
B^{\mathsf T}$ onto the column space of $B$.

\begin{lemma}[A half-sample bound]\label{rec:half}
Fix a list indexed by $I$, and include its indices in $S$ independently
with probability $1/2$. Then
\[
 \mathbb E_S\sum_{i\in I}\sigma_{L_S}(b_i)\le3n.
\]
\end{lemma}

\begin{proof}
Fix $i$ and condition on all membership decisions other than that of $i$.
Write $M=L_{S\setminus\{i\}}$ and $t=R_M(b_i)$.
If $t$ is finite, the rank-one inverse identity on $\operatorname{im}M$
gives
\[
 R_{M+b_i b_i^{\mathsf T}}(b_i)=\frac{t}{1+t}.
\]
If $b_i\notin\operatorname{im}M$, the new resistance is $1$.
For example, in the minimum-norm characterization of leverage, the
component of $b_i$ orthogonal to $\operatorname{im}M$ forces its own
coefficient to equal $1$; taking all other coefficients zero attains
this norm. Set $\ell=t/(1+t)$ for finite $t$, and $\ell=1$ for $t=+\infty$.
In the following indicator-weighted expression the value is defined to
be zero when $i\notin S$, avoiding any product $0\cdot(+\infty)$.
Then
\begin{align*}
 \mathbb E[\sigma_{L_S}(b_i)\mid S\setminus\{i\}]
 &=\tfrac12\min\{1,t\}+\tfrac12\ell\\
 &\le\tfrac32\ell
 =3\mathbb E[\mathbf1_{\{i\in S\}}R_{L_S}(b_i)
                  \mid S\setminus\{i\}].
\end{align*}
The inequality uses $\min\{1,t\}\le2t/(1+t)$, with the same conclusion
at $t=0$ and $t=+\infty$. Summing, removing the conditioning, and
applying~\eqref{rec:rank-identity} proves the claim.
\end{proof}

\begin{lemma}[Scores from an approximate half-sample]\label{rec:scores}
Suppose $S\subseteq I$ and an explicit matrix $H$ satisfies
\[
 \tfrac12L_S\preceq H\preceq\tfrac32L_S.
\]
Define $u_i=\min\{1,2R_H(b_i)\}$, taking $u_i=1$ outside
$\operatorname{im}H$. For every $i\in I$,
\[
 R_{L_I}(b_i)\le u_i\le4\sigma_{L_S}(b_i).
\]
\end{lemma}

\begin{proof}
The sandwich gives $\ker H=\ker L_S$. On their common range, inverse
order gives
\[
 \tfrac23L_S^\dagger\preceq H^\dagger\preceq2L_S^\dagger.
\]
Outside this range, $u_i=\sigma_{L_S}(b_i)=1$ and
$R_{L_I}(b_i)\le1$, proving both assertions.
Inside the range,
$\tfrac43R_{L_S}(b_i)\le2R_H(b_i)\le4R_{L_S}(b_i)$.
Since $L_S\preceq L_I$, resistance cannot increase upon passing from
$L_S$ to $L_I$ for a vector in $\operatorname{im}L_S$. This also follows
directly from the variational formula
\[
 R_M(b)=\sup_x\{2b^{\mathsf T}x-x^{\mathsf T}Mx\},
 \qquad b\in\operatorname{im}M.
\]
Taking the minimum with $1$, and using $R_{L_I}(b_i)\le1$, proves the result.
\end{proof}

\begin{lemma}[Sampling with upper leverage scores]\label{rec:sampling}
Suppose $u_i\in[0,1]$ and $u_i\ge R_{L_I}(b_i)$ for all $i\in I$.
Fix $0<a\le1/2$ and $0<\beta<1$, and put
\[
 \alpha=\left\lceil8a^{-2}\left\lceil\log_2(2n/\beta)\right\rceil
                 \right\rceil,
 \qquad p_i=\min\{1,\alpha u_i\}.
\]
Let $q_i\in[p_i,1]$. Independently retain each index with probability
$q_i$, omit indices with $q_i=0$, and form
\[
 \widehat L=\sum_{i\in I:q_i>0}\frac{\xi_i}{q_i}b_i b_i^{\mathsf T},
 \qquad \xi_i\sim\operatorname{Bernoulli}(q_i).
\]
With probability at least $1-\beta$,
$(1-a)L_I\preceq\widehat L\preceq(1+a)L_I$.
\end{lemma}

\begin{proof}
The assertion is immediate if $L_I=0$. Otherwise work on
$\operatorname{im}L_I$, put $v_i=(L_I^\dagger)^{1/2}b_i$, and let $P$
be the orthogonal projection onto that space. Thus
$\sum_i v_i v_i^{\mathsf T}=P$ and
$\|v_i\|^2=R_{L_I}(b_i)\le u_i$.
Indices with $q_i=1$ have zero centered contribution. For other positive
probabilities $q_i\ge\alpha u_i$, so the independent mean-zero matrices
$Y_i=(\xi_i/q_i-1)v_i v_i^{\mathsf T}$ satisfy
\[
 \|Y_i\|\le\frac{\|v_i\|^2}{q_i}\le\alpha^{-1},
 \qquad
 \sum_i\mathbb EY_i^2
 =\sum_i(q_i^{-1}-1)\|v_i\|^2v_i v_i^{\mathsf T}
 \preceq\alpha^{-1}P.
\]
Zero-probability indices have $u_i=0$ and hence $v_i=0$, so contribute
nothing. Self-adjoint matrix Bernstein~\cite[Theorem 6.1]{Tropp12} gives
\[
 \Pr\!\left(\left\|\sum_iY_i\right\|>a\right)
 \le2n\exp\!\left(-\frac{a^2}{2/\alpha+2a/(3\alpha)}\right)
 \le\beta.
\]
The last inequality follows from the chosen $\alpha$ and $a\le1/2$.
Conjugating on the range proves the sandwich there. All summands vanish
on $\ker L_I$, so it holds on the entire space.
\end{proof}

\begin{proof}[Proof of Lemma~\ref{rec:main}]
Set
\[
 \ell=\left\lceil\log_2(4N/\delta)\right\rceil,
 \qquad \beta=\frac{\delta}{10(\ell+1)}.
\]
Generate independent Bernoulli-half mask tables, defining nested known
index sets $I_0=[N],I_1,\ldots,I_\ell$, with $I_{j+1}$ a half-sample of
$I_j$. These are masks on indices; generating and looking up the masks
does not query a record. Since an index survives with probability
$2^{-\ell}$,
\[
 \Pr(I_\ell\ne\varnothing)\le N2^{-\ell}\le\delta/4.
\]
Abort if $I_\ell\ne\varnothing$. Otherwise $H_\ell=0$ represents the
empty sublist exactly. No assumption that query-domain sizes shrink
geometrically is needed.

For $j=\ell-1,\ldots,0$, use the explicit child $H_{j+1}$ to construct
$H_j$ from the actual records indexed by $I_j$. Request accuracy
$a_j=1/2$ for $j\ge1$ and $a_0=\varepsilon$.
A correct child satisfies
$\tfrac12L_{I_{j+1}}\preceq H_{j+1}\preceq\tfrac32L_{I_{j+1}}$,
also at the bottom. Lemma~\ref{rec:half} and Markov's inequality give
\begin{equation}\label{rec:score-event}
 \sum_{i\in I_j}\sigma_{L_{I_{j+1}}}(b_i)\le3n/\beta
\end{equation}
except on an event of probability at most $\beta$, conditional on any
fixed $I_j$ and hence also unconditionally. Whenever the displayed bound
and child correctness both hold, Lemma~\ref{rec:scores} implies
$\sum_{i\in I_j}u_i\le12n/\beta$.
The bound on the bad event~\eqref{rec:score-event} is not conditioned on
child success or on $I_\ell=\varnothing$. Those events can be correlated
with the half-sample. We will union-bound the half-sample bad events
unconditionally.

At this level let
\[
 \alpha_j=\left\lceil8a_j^{-2}
                  \left\lceil\log_2(2n/\beta)\right\rceil\right\rceil,
 \qquad b=\left\lceil\log_2\max\{2,N\}\right\rceil.
\]
For a queried record compute $p_i=\min\{1,\alpha_j u_i\}$ and round
upward to a multiple $q_i$ of $2^{-b}$, keeping $q_i=0$ when $p_i=0$
and $q_i=1$ when $p_i=1$. Since $2^{-b}\le1/N$, the total increase
in sampling probabilities is at most $1$. On the preceding score event
and with a correct child,
\[
 \sum_{i\in I_j}q_i\le\mu_j,
 \qquad \mu_j=12\alpha_j n/\beta+1.
\]
Choose fresh independent uniform integers in $\{0,\ldots,2^b-1\}$
for every index at this level. Comparing the integer for index $i$ with
$2^bq_i$ implements its Bernoulli decision exactly. The child Laplacian
has rational entries, and its pseudoinverse and image membership can be
computed by exact rational linear algebra. Consequently every score
comparison, sampling probability, and retained weight $1/q_i$ is finite
and exact. Rational accuracy and failure parameters suffice for all
applications below; for other specified accuracies one can use smaller
rational parameters within a constant factor.

Impose the hard output cap
\[
 B_j=\min\left\{N,
       \left\lceil2\mu_j+6\left\lceil\log_2(1/\beta)\right\rceil
                         \right\rceil\right\}.
\]
Conditional on the masks and acquired child data satisfying the preceding
good events, fresh coins remain independent. Scalar Chernoff bounds show
that the probability of more than $B_j$ retained records is at most
$\beta$, and is zero when $B_j=N$.
Conditional on a correct child, Lemma~\ref{rec:scores} supplies upper
leverage scores for the actual current list, so Lemma~\ref{rec:sampling}
gives another failure probability at most $\beta$ for its spectral
sandwich. This conclusion does not require conditioning on the realized
sample size.

To acquire the sample, enumerate marked indices in the original domain
$[N]$. An index is marked precisely when it is in $I_j$ and its fresh
integer is below the threshold computed from its record. One coherent
predicate evaluation uses a constant number of original record queries,
including uncomputation. All masks, coins, and child-matrix data are
explicit known tables. Capped quantum enumeration
\cite[Theorem 13]{AL21} with marked-count promise $B_j$ uses
$\OO(\sqrt{N(B_j+1)})$ evaluations and can have failure probability
at most $\beta$. Impose this fixed evaluation budget even if the promise
fails. Allow at most $B_j$ returned indices and at most $B_j$ further
record queries to retrieve them. Thus failure of any earlier step cannot
increase the query or retrieval budgets.

If $B_j=N$, read the current sublist and use its exact Laplacian as
$H_j$ instead. Otherwise, reject reported indices outside $I_j$;
after capped retrieval recompute their probabilities and marking tests.
Reject a zero-probability index or a failed marking test before forming
$1/q_i$, and deduplicate indices. The resulting explicitly weighted
sample is $H_j$. An unsuccessful enumeration need not be recognizable
in every case; it is included in the failure probability. The imposed
caps hold irrespective of whether an unsuccessful output is recognized.

For completeness, first exclude $I_\ell\ne\varnothing$ and all
half-sample bad events~\eqref{rec:score-event} using their unconditional
probability bounds. At each level, on a successful history, an oversized
sample, a failed sampled spectral sandwich, or an unsuccessful
enumeration each has probability at most $\beta$ on the appropriate
preceding data. The sample-size and spectral failures may be dependent;
their individual conditional bounds suffice for a union bound. Applying
this reasoning to the first remaining failure in bottom-up order gives
total failure probability at most
\[
 \delta/4+4\ell\beta<\delta.
\]
There is no multiplicative accumulation of spectral errors: $H_j$ is
sampled from the actual rows in $I_j$ and approximates $L_{I_j}$ directly.
The child only supplies sampling probabilities. Thus the root accuracy
is $\varepsilon$.

Every level uses a query domain of size at most $N$, with enumeration
and retrieval budget
\[
 \OO\!\left(\frac{\sqrt{Nn}}{a_j\sqrt\beta}\right).
\]
When $B_j<N$, retrieval is absorbed by $B_j\le\sqrt{NB_j}$.
When $B_j=N$, its definition implies $N=O(\alpha_j n/\beta)$, so reading
the whole sublist is absorbed by the same displayed bound.
Summing over the fixed $\ell$ levels, using $a_j=1/2$ except at the root,
and substituting $\beta=\delta/[10(\ell+1)]$ gives
$\OO(\sqrt{Nn}/(\varepsilon\sqrt\delta))$.
All loops and oracle subroutines have fixed budgets, including on failed
sampling outcomes. Choose the read-all alternative from the known
parameters whenever it is cheaper.
\end{proof}

\bibliographystyle{plain}
\bibliography{references}

@inproceedings{RSW18,
 author={Aviad Rubinstein and Tselil Schramm and S. Matthew Weinberg},
 title={Computing Exact Minimum Cuts Without Knowing the Graph},
 booktitle={9th Innovations in Theoretical Computer Science Conference (ITCS 2018)},
 series={LIPIcs}, volume={94}, pages={39:1--39:16}, year={2018},
 doi={10.4230/LIPIcs.ITCS.2018.39},
 note={Full version: arXiv:1711.03165; Lemma 2.6 and Section 2.3},
 url={https://arxiv.org/abs/1711.03165}
}

@inproceedings{AL21,
 author={Simon Apers and Troy Lee},
 title={Quantum Complexity of Minimum Cut},
 booktitle={36th Computational Complexity Conference (CCC 2021)},
 series={LIPIcs}, volume={200}, pages={28:1--28:33}, year={2021},
 doi={10.4230/LIPIcs.CCC.2021.28},
 note={Full version: arXiv:2011.09823v3; structural Lemma 2},
 url={https://arxiv.org/abs/2011.09823}
}

@misc{AAL24,
 author={Simon Apers and Arinta Auza and Troy Lee},
 title={A sublinear query quantum algorithm for {$s$-$t$} minimum cut on dense simple graphs},
 year={2024}, howpublished={arXiv:2110.15587v2},
 note={Revised 5 February 2024; query-complexity guarantee},
 url={https://arxiv.org/abs/2110.15587}
}

@article{Ambainis02,
 author={Andris Ambainis}, title={Quantum lower bounds by quantum arguments},
 journal={Journal of Computer and System Sciences},
 volume={64}, number={4}, pages={750--767}, year={2002},
 doi={10.1006/jcss.2002.1826},
 note={Preliminary version STOC 2000; arXiv:quant-ph/0002066},
 url={https://arxiv.org/abs/quant-ph/0002066}
}

@inproceedings{NayakWu99,
 author={Ashwin Nayak and Felix Wu},
 title={The Quantum Query Complexity of Approximating the Median and Related Statistics},
 booktitle={Proceedings of the 31st Annual ACM Symposium on Theory of Computing},
 pages={384--393}, year={1999}, doi={10.1145/301250.301349},
 note={arXiv:quant-ph/9804066}, url={https://arxiv.org/abs/quant-ph/9804066}
}

@inproceedings{AS06,
 author={Andris Ambainis and Robert {\v{S}palek}},
 title={Quantum Algorithms for Matching and Network Flows},
 booktitle={STACS 2006}, series={Lecture Notes in Computer Science},
 volume={3884}, pages={172--183}, year={2006}, doi={10.1007/11672142_13},
 note={arXiv:quant-ph/0508205; Theorem 7},
 url={https://arxiv.org/abs/quant-ph/0508205}
}

@inproceedings{BGMP21,
 author={Arijit Bishnu and Arijit Ghosh and Gopinath Mishra and Manaswi Paraashar},
 title={Query Complexity of Global Minimum Cut},
 booktitle={Approximation, Randomization, and Combinatorial Optimization. Algorithms and Techniques (APPROX/RANDOM 2021)},
 series={LIPIcs}, volume={207}, pages={6:1--6:15}, year={2021},
 doi={10.4230/LIPIcs.APPROX/RANDOM.2021.6},
 url={https://arxiv.org/abs/2007.09202}
}

@inproceedings{Li19,
 author={Jason Li}, title={Faster Minimum {$k$}-cut of a Simple Graph},
 booktitle={60th IEEE Symposium on Foundations of Computer Science (FOCS 2019)},
 year={2019}, pages={1056--1077}, doi={10.1109/FOCS.2019.00068},
 note={arXiv:1910.02665; Theorem 3.2},
 url={https://arxiv.org/abs/1910.02665}
}

@article{Wahlstrom22,
 author={Magnus Wahlstr{\"o}m},
 title={Quasipolynomial Multicut-mimicking Networks and Kernels for Multiway Cut Problems},
 journal={ACM Transactions on Algorithms}, volume={18}, number={2},
 pages={15:1--15:19}, year={2022}, doi={10.1145/3501304},
 note={Preliminary version ICALP 2020; arXiv:2002.08825},
 url={https://arxiv.org/abs/2002.08825}
}

@inproceedings{ALLS24,
 author={Aditya Anand and Euiwoong Lee and Jason Li and Thatchaphol Saranurak},
 title={Approximating Small Sparse Cuts},
 booktitle={Proceedings of the 56th Annual ACM Symposium on Theory of Computing},
 year={2024}, note={arXiv:2403.08983},
 url={https://arxiv.org/abs/2403.08983}
}

@inproceedings{IKPRS24,
 author={Tanmay Inamdar and Madhumita Kundu and Pekka Parviainen and M. S. Ramanujan and Saket Saurabh},
 title={Exponential-Time Approximation Schemes via Compression},
 booktitle={15th Innovations in Theoretical Computer Science Conference (ITCS 2024)},
 series={LIPIcs}, volume={287}, pages={64:1--64:22}, year={2024},
 doi={10.4230/LIPIcs.ITCS.2024.64}
}

@misc{Cohen14,
 author={Michael B. Cohen and Yin Tat Lee and Cameron Musco and Christopher Musco and Richard Peng and Aaron Sidford},
 title={Uniform Sampling for Matrix Approximation},
 year={2014}, howpublished={arXiv:1408.5099},
 url={https://arxiv.org/abs/1408.5099}
}

@article{Tropp12,
 author={Joel A. Tropp},
 title={User-Friendly Tail Bounds for Sums of Random Matrices},
 journal={Foundations of Computational Mathematics},
 volume={12}, number={4}, pages={389--434}, year={2012},
 doi={10.1007/s10208-011-9099-z},
 note={Theorem 6.1; arXiv:1004.4389},
 url={https://arxiv.org/abs/1004.4389}
}
\end{document}